\documentclass[%
aps,
prd,
twocolumn,
preprintnumbers,
nofootinbib,
amsmath,amssymb,
]{revtex4-2}

\usepackage{style}
\usepackage{aas_macros}

\begin{document}

\preprint{DESY-23-192}

\title{Stochastic Ultralight Dark Matter Fluctuations in Pulsar Timing Arrays}

\author{Hyungjin Kim\,\orcidlink{0000-0002-8843-7690}}
\affiliation{Deutsches Elektronen-Synchrotron DESY, Notkestr. 85, 22607 Hamburg, Germany}

\author{Andrea Mitridate\,\orcidlink{0000-0003-2898-5844}}
\affiliation{Deutsches Elektronen-Synchrotron DESY, Notkestr. 85, 22607 Hamburg, Germany}

\begin{abstract}
Metric perturbations induced by ultralight dark matter (ULDM) fields have long been identified as a potential target for pulsar timing array (PTA) observations. Previous works have focused on the coherent oscillation of metric perturbations at the characteristic frequency set by the ULDM mass. In this work, we show that ULDM fields source low-frequency stochastic metric fluctuations and that these low-frequency fluctuations can produce distinctive detectable signals in PTA data. Using the NANOGrav 12.5-year data set and synthetic data sets mimicking present and future PTA capabilities, we show that the current and future PTA observations provide the strongest probe of ULDM density within the solar system for masses in the range of $10^{-18}\;{\rm eV}-10^{-16}\;{\rm eV}$.
\end{abstract}

\maketitle
\tableofcontents

\section{Introduction}
Millisecond pulsars are one of nature’s most precise clocks, with a frequency stability that rivals one of human-made atomic clocks. By monitoring the ticking of a net of these cosmic clocks, pulsar timing arrays (PTAs) have achieved unparalleled sensitivity to spacetime perturbations in the ${\rm nHz}$ band. By leveraging this remarkable sensitivity, several PTA collaborations have recently found evidence for the presence of a gravitational wave background permeating our galaxy~\cite{NANOGrav:2023gor, Antoniadis:2023rey, Reardon:2023gzh, Xu:2023wog}.

While the primary objective of PTAs is the detection of gravitational waves (GWs), we can also leverage their sensitivity to search for spacetime fluctuations produced by different sources. A notable example is the metric fluctuations induced by ultralight dark matter---a bosonic dark matter candidate with a mass smaller than a few eV. Such light bosonic dark matter candidates behave like a classical wave, coherently oscillating at a frequency set by their mass, $m_\phi$.
These oscillations will cause time-dependent metric fluctuations at a frequency $\omega =2m_\phi$, making ULDM candidates in the mass window $10^{-23}\eV-10^{-20}\eV$ potentially detectable by PTAs, as first demonstrated by Khmelnitsky and Rubakov~\cite{Khmelnitsky:2013lxt}.
In addition, if ULDM couples to the Standard Model particles non-gravitationally, it will affect the spin of pulsars as well as the frequency of atomic clocks used for pulsar timing measurements, leading to another monochromatic signal in the timing observation~\cite{Graham:2015ifn, Kaplan:2022lmz, NANOGrav:2023hvm}.

\begin{figure}
    \centering
    \includegraphics{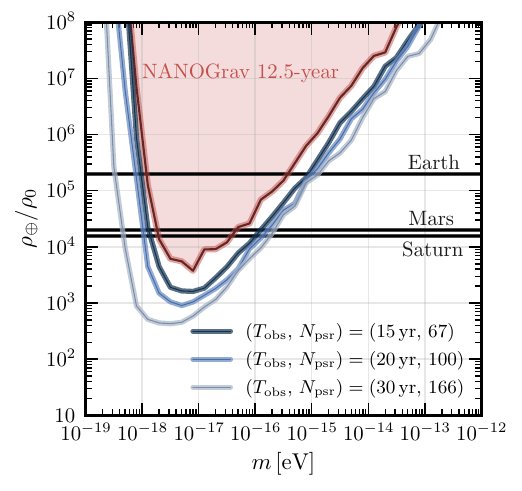}
    \caption{The red line shows the 95\% constraints on the dark matter density near the solar system derived by analyzing the NANOGrav 12.5-year data set. The blue lines show projections derived by analyzing mock data sets with  
    $N_{\rm psr} =66$ pulsar and a total observing time of $T_{\rm obs} = 15\,{\rm yr}$ (dark blue), $N_{\rm psr} = 100$ pulsars and $T_{\rm obs} = 20 \, {\rm yr}$ observing time (medium blue), and $N_{\rm psr} = 166$ pulsars and $T_{\rm obs} =30 \,{\rm yr}$ observing time (light blue). These constraints are compared with the ones derived from solar system ephemerides (black lines) at Earth, Mars, and Saturn orbit. All the constraints are normalized to $\rho_0=0.4\,{\rm GeV/cm}^3.$ }
    \label{fig:summary}
\end{figure}

However, ULDM density fluctuations exhibit much richer temporal and spectral behaviors than a simple coherent oscillation. In a recent work~\cite{Kim:2023pkx}, it was shown that, in addition to the coherent oscillation at $\omega = 2m_\phi$, ULDM fields exhibit low-frequency stochastic fluctuations at $\omega \lesssim m_\phi \sigma^2$, where $\sigma \simeq 160\,{\rm km/sec}$ is the velocity dispersion of the virialized dark matter halo. These low-frequency fluctuations are intricately related to the order-one density fluctuations within the ULDM halo observed in numerical simulations (e.g. Ref.~\cite{Schive:2014dra}) over characteristic time and length scales given by the coherence time, $\tau = 1/m_\phi \sigma^2$, and the coherence length, $\lambda = 1/m_\phi \sigma$. Implications of these low-frequency stochastic fluctuations in PTA observations have not yet been investigated and will be the primary focus of this work. 

We can intuitively understand these low-frequency fluctuations in terms of {\it quasiparticles}, whose size is set by the coherence length and whose mass is given by $m_{\rm eff} =\pi^{3/2} \rho_{\scriptscriptstyle\rm DM} \lambda^3$, with $\rho_{\scriptscriptstyle\rm DM}$ being the ULDM density~\cite{Hui:2016ltb, Bar-Or:2018pxz}.
In Ref.~\cite{Kim:2023pkx}, it was shown that the gravitational interaction between these quasiparticles and test masses in GW interferometers inevitably introduces stochastic signals, potentially detectable by existing and planned future GW detectors. It was also found that interferometers with longer baselines are better suited to look for such stochastic signals---since they are sensitive to larger quasiparticles which can impart larger accelerations on the test masses. In the same study, it was also highlighted that more than two detectors would be required to distinguish stochastic ULDM signals from detector noise. For these reasons, PTAs arise as a natural candidate to look for such stochastic ULDM fluctuations. Thanks to the $\mathcal{O}(\rm{kpc})$ distance between pulsars and the Earth, PTAs can be perturbed by larger quasiparticles compared to test masses in Earth-based interferometers. Moreover, correlations of the ULDM signal across the pulsars in the array provide a natural means to distinguish ULDM signals from detector-specific noise.

In the quasiparticle picture, one can easily estimate the perturbation to the TOAs induced by stochastic ULDM fluctuations. For PTAs with an observing baseline, $T_{\rm obs}$, larger than the ULDM coherence times, a typical TOA fluctuation is approximately given by:
\begin{align}\label{eq:signal_estimate}
    \delta t
    &\simeq c \, 
    (\bar a \tau^2) 
    \Big( \frac{T_{\rm obs}}{\tau} \Big)^{\frac32}
    \nonumber\\
    &\sim 10^{-10}\,{\rm sec}
    \left(\frac{10^{-17}\,{\rm eV}}{m_\phi}\right)^{\frac32}
    \left(\frac{\rho_\oplus}{0.4\,{\rm GeV}/{\rm cm}^3}\right)
\end{align}
where $\bar a = G m_{\rm eff}/\lambda^2$ is a typical acceleration on Earth produced by ULDM quasiparticles, $c \sim {\cal O}(10^{-2} \, \textrm{--} \, 10^{-1})$ is a numerical coefficient, and $\rho_\oplus$ is an average ULDM density around the solar system. This expression is valid for $T_{\rm obs} / \tau \gg 1$, and we choose $T_{\rm obs}=30\,{\rm yr}$ for the estimation. In the expression, the quantity in the first parenthesis ($\bar a \tau^2$) can be understood as a typical displacement of Earth due to quasiparticles over one coherence time scale and the quantity $(T_{\rm obs}/\tau)^{3/2}$ can be understood as a result of random walk motion of the relative velocity of Earth-pulsar system induced by ULDM quasiparticles. 

A naive comparison of this pulsar timing fluctuation with typical TOA errors $\sigma_{\rm TOA} \sim {\cal O}(10^{-7}) \,{\rm sec}$ already suggests that the current PTA might probe average solar-system ULDM densities of $\rho_\oplus/\rho_0 \sim 10^3$ at $m_\phi =10^{-17}\eV$, where $\rho_0\simeq0.4\;{\rm GeV}/{\rm cm}^3$ represents typical local dark matter density. It is important to note that this local density, $\rho_0$, is measured on volumes of at least $\sim (10^2\;{\rm pc})^3$ surrounding the solar system (see for example Refs.~\cite{Read:2014qva, deSalas:2020hbh}), while there is currently no direct measurement of dark matter density within and around the solar system. This naive estimate suggests that PTAs might provide one of the strongest probes of ULDM density near the solar system. The main objective of this work is to sharpen the above estimate by carefully characterizing the effects of stochastic ULDM fluctuations in current and future PTA observations.

We summarize our main results in Figure~\ref{fig:summary} where we show the constraints and projections on the ULDM density near the solar system derived by analyzing the NANOGrav 12.5-year data set~\cite{NANOGrav:2020gpb}, and a collection of simulated data sets mimicking the capabilities of future observations (see Section~\ref{subsec:mock} for details on the generation of the mock data). A null result from the analysis of NANOGrav 12.5-year data sets the strongest upper limit on ULDM density near the solar system as $\rho_\oplus/\rho_0 \lesssim 4\times 10^3$, improving the existing constraints from solar system ephemerides $\rho_\oplus /\rho_0 \lesssim 2\times 10^4$~\cite{Pitjev:2013sfa} in the mass range $10^{-18}\;{\rm eV}-10^{-16}\;{\rm eV}$. The analyses with simulated data sets show that current constraints could be improved by an order of magnitude or more with a larger number of pulsars with a longer observation period, which could be easily achieved in future PTAs. 

The paper is organized as follows. In section~\ref{sec:ULDM}, we discuss the statistical property of the ULDM field. We compute the timing residual power spectrum induced by stochastic ULDM fluctuations and derive their correlation pattern. In section~\ref{sec:analysis}, we detail our analysis: we discuss the data analysis tools used in our analysis, and how we generated the mock data sets. In section~\ref{sec:result}, we present the result from our analysis of the mock pulsar timing data in the context of ultralight dark matter. We conclude in section~\ref{sec:conclusion}. 

\section{Stochastic ULDM signal}\label{sec:ULDM}
Pulsar timing arrays track the TOAs of radio pulses emitted by a collection of galactic millisecond pulsars. These TOAs are then compared with the predictions of a pulsar timing model. The discrepancies between these predictions and the observed TOAs define the \emph{timing residuals}, $\delta t$. In this section, we will discuss the imprint of stochastic ULDM fluctuations on these timing residuals. 

We parametrize the metric perturbations induced by fluctuations of the ULDM field as 
\begin{equation}
    ds^2=\Big[1+2\Phi(t,\vec{x})\Big]dt^2
    -\Big[1-2\Psi(t,\vec{x})\Big]dx^2\,.
\end{equation}
These metric perturbations can influence the TOAs and leave an imprint in the timing residuals in several different ways:

\medskip\noindent\textbullet~\emph{Doppler effect:} The force exerted by ULDM fluctuations induces an acceleration of both Earth and pulsars. This results in perturbations of the perceived pulsar period due to the induced relative motion between the observer (Earth) and the emitters (pulsars). We can write these perturbations in terms of the scalar potential $\Phi$ as:
\begin{equation}\label{eq:doppler_signal}
    \frac{\delta\nu_a(t)}{\nu_a}=\hat{\vec{n}}_a\cdot\int^t dt'\;\nabla\Big[\Phi(t'-d_a,\vec{x}_p)-\Phi(t',\vec{x}_e)\Big]\,,
\end{equation}
where $\nu_a$ is the perceived pulsar rotational frequency for the $a$-th pulsar in the array, $\hat{\vec{n}}_a$ is a unit vector pointing from Earth to pulsar, $d_a$ is Earth-pulsar distance, and $\vec{x_e}$ and $\vec{x}_p$ represent Earth and pulsar position respectively.

\medskip\noindent\textbullet~\emph{Other effects:} ULDM fluctuations source metric fluctuation between Earth and pulsars, thereby perturbing the light travel time along the line-of-sight. This effect is often called as Shapiro delay. Additionally, metric fluctuations at the position of the Earth and pulsar change the measurements of TOAs and pulsar period as both are measured by the proper time of observer at the Earth and pulsar, respectively. This effect is called Einstein delay. Combined, they induce a shift in the perceived rotational pulsar period as
\begin{equation}
    \begin{split}
        \frac{\delta\nu_a(t)}{\nu_a}=& \Psi(\vec{x}_e,t)-\Psi(\vec{x}_p, t-d_a)\\
        -&\hat{\vec{n}}_a\cdot\int_{t-d_a}^t dt'\;\nabla\Big[\Phi(t',\vec{x}(t'))+\Psi(t',\vec{x}(t'))\Big]
    \end{split}
\end{equation}
where $\vec{x}(t')=\vec{x}_e+\hat{\vec{n}}_a(t-t')$, i.e. the integral must be performed along the photon path.  

\medskip 
Timing residuals can be computed from fluctuations in the arrival frequency as
\begin{equation}\label{eq:nu_to_dt}
    \delta t_a(t)=\int^t\frac{\delta\nu_a(t')}{\nu_a}dt'\,.
\end{equation}
In this work, we focus on the Doppler signal given in Eq.~\eqref{eq:doppler_signal} as the Shapiro and Einstein effects are suppressed by an additional factor of dark matter velocity. See Appendix~\ref{app:timing_residuals} for more details.

\subsection{ULDM Signal Power Spectrum}
The goal of this section is to relate the PTA signals discussed in the previous section with ULDM density fluctuations, $\delta\rho$, defined as
\begin{equation}
    \delta\rho\equiv\rho-\rho_\oplus\,,
\end{equation}
where, as already mentioned before, $\rho_\oplus$ is the average ULDM density within the solar system. 
Specifically, since we are interested in stochastic ULDM fluctuations we want to relate the power spectrum of ULDM density fluctuations to the power spectrum of PTA timing residuals. The first of these two quantities is defined as 
\begin{equation}\label{eq:rho_ps}
    \langle \widetilde{\delta\rho}(k) \widetilde{\delta\rho}^*(k') \rangle = (2\pi)^4 \delta^{(4)}(k-k') P_{\delta\rho}(k)\,.
\end{equation}
where $\tilde\delta\rho(k)$ denotes the Fourier components of the ULDM density fluctuations, and the angle brackets denote the ensemble average over all possible realizations of the ULDM field.

Following the procedure outlined in Refs.~~\cite{Kim:2021yyo, Kim:2023pkx, Kim:2023pvt}, we can show that the power spectrum for ULDM density fluctuations can be expressed in terms of the DM phase space distribution, $f(\vec{p})$, as (see Appendix~\ref{app:fluctuation} for more details):
\begin{equation} \label{eq:density_fluctuation_PS}
    \begin{split}
    P_{\delta\rho}
    \approx 
    m_\phi^2 
    \int\frac{d^3p_1}{(2\pi)^3}
    \frac{d^3p_2}{(2\pi)^3}
    f(\vec p_1) f(\vec p_2) (2\pi)^4 \delta^4(k - p_1 + p_2),
    \end{split}
\end{equation}
where we have ignored the modes at $\omega = \pm (\omega_1 + \omega_2) \simeq \pm 2m_\phi$, which represents the coherently oscillating modes not of interest for the analysis discussed in this work. Using Poisson equation,  $\nabla^2 \Phi = 4\pi G \delta\rho$,\footnote{For the low-frequency stochastic fluctuations, one can approximate $\Psi \approx \Phi$~\cite{Kim:2023pkx}.} the power spectrum of ULDM density fluctuations can be easily related to the power spectrum of the scalar potential fluctuations, $P_\Phi$, as
\begin{equation}\label{eq:poisson}
    P_{\Phi}(k) = \frac{(4\pi G)^2}{k^4} P_{\delta \rho} (k)\,,
\end{equation}
where $G$ is the gravitational constant, and $P_\Phi$, is defined analogously to the density power spectrum in Eq.~\eqref{eq:rho_ps}.

The power spectrum for scalar perturbations can also be related to the timing residuals covariance matrix, defined as:
\begin{equation}\label{eq:dtdt}
    \langle \delta t_{ai} \delta t_{bj} \rangle 
    \supset\Gamma_{ab}^{\scriptscriptstyle\rm DM}  \int df \cos[2\pi f(t_{ai}-t_{bj})]\, S^{\scriptscriptstyle\rm DM}(f)
\end{equation}
where $\Gamma_{ab}^{\scriptscriptstyle\rm DM}$ is the overlap reduction function (ORF) characterizing the correlations among timing residual of different pulsars, $S^{\scriptscriptstyle\rm DM}(f)$ is the timing residual power spectrum, and the limits of the integration are determined by the inverse of the observational cadence and the total observation period. Here $i$ and $j$ index TOAs, and $a$ and $b$ index pulsars. By using Eq.~\eqref{eq:doppler_signal} together with Eqs.~\eqref{eq:nu_to_dt} and \eqref{eq:dtdt}, we can relate $S^{\scriptscriptstyle\rm DM}_{ab}\equiv\Gamma_{ab}^{\scriptscriptstyle\rm DM}S^{\scriptscriptstyle\rm DM}(f)$ with fluctuations of the gravitational potential:
\begin{equation}\label{eq:PS_dt}
    S^{\scriptscriptstyle\rm DM}_{ab}(f)=
     \frac{2}{(2\pi f)^4}
    \int \frac{d^3k}{(2\pi)^3} 
    (\hat n_a \cdot \vec k)
    (\hat n_b \cdot \vec k)
    U_a U_b^*
    P_{\Phi}(f,\vec k),
\end{equation}
where $U_a = 1 - \exp\big( 2\pi i f d_a [ 1 + (k/\omega) \hat k \cdot \hat n_a] \big)$. Given that $2\pi f d_a \gg1$ for the frequency range of PTA observations and typical Earth-pulsar distances, we can approximate $U_a U_b^* \approx 1 + \delta_{ab}$.

Finally, using the relation in Eq.~\eqref{eq:poisson}, we can rewrite Eq.~\eqref{eq:PS_dt} as 
\begin{equation}
    S^{\scriptscriptstyle\rm DM}_{ab}(f)=\frac{2 G^2}{\pi^2 f^4}
    \int \frac{d^3k}{(2\pi)^3} 
    (\hat n_a \cdot \vec k)
    (\hat n_b \cdot \vec k)
    U_a U_b^*
    \frac{P_{\delta\rho}(f,\vec k)}{k^4}.
\end{equation}
In the two following sections, we will use this expression to derive the timing residual signal for two specific forms of the DM phase space distribution. 

\begin{figure*}
    \centering
    \includegraphics{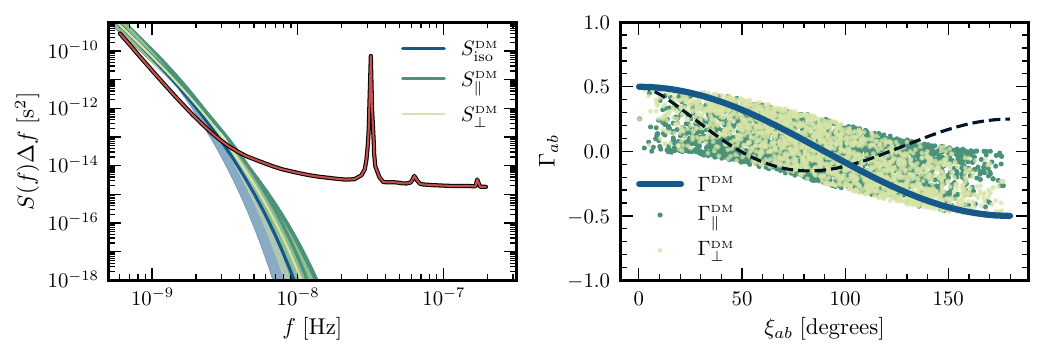}
    \caption{\textbf{Left:} Comparison between the sensitivity curve for the NANOGrav 11-year data set~\cite{Hazboun:2019vhv} (red), the characteristic strain for the ULDM signal in the isotropic limit (blue), and the two components of the ULDM strain in the anisotropic limit (light and dark green). The shaded bands are derived by varying the value of the velocity dispersion between $\sigma=150\;{\rm km/s}$ and $\sigma=250\;{\rm km/s}$. In this plot we have chosen $\rho_\oplus/\rho_0=10^4$, and $m_\phi=10^{-17}\;{\rm eV}$. \textbf{Right:} The blue line shows the overlap reduction function for the ULDM signal in the isotropic limit. Light and dark green dots represent the correlations amongst the pulsars in our 15-year mock data set for the ULDM signal in the anisotropic limit. The dashed black line represents the Hellings \& Downs correlation pattern.}
    \label{fig:spectra}
\end{figure*}

\subsection{Isotropic Distribution}
If we ignore the motion of the solar system with respect to the galactic center, the DM has an isotropic velocity distribution that can be written as
\begin{equation}
    f(\vec v) = \frac{\rho_\oplus/m_\phi}{(2\pi\sigma^2)^{3/2}}
    \exp\left[ - \frac{v^2}{2\sigma^2} \right].
\end{equation}
In this case, we can derive simple analytic expressions for the timing-residual power spectrum and its ORF (see Appendix~\ref{app:timing_residuals} for a derivation):
\begin{align}
    \Gamma_{ab}^{\scriptscriptstyle\rm DM}
    &= 
    \frac{1}{2} \big( \delta_{ab} + \hat n_a \cdot \hat n_b \big),
    \\
    S^{\scriptscriptstyle\rm DM}(f)
    &=
    \frac{\bar a^2 \tau}{(2\pi f)^4} 
    \left[
    \frac{64}{3\pi} 
    K_0( \omega\tau )\right]. \label{eq:P_uldm_iso}
\end{align}
Here $\tau = 1 / m_\phi \sigma^2$ is the coherence time, $\bar a = G m_{\rm eff}/ \lambda^2$ is the typical acceleration due to ULDM quasiparticles, $m_{\rm eff} = \pi^{3/2}\rho_\oplus \lambda^3$ is the quasiparticle mass, $\lambda=1/m_\phi \sigma$ is the wavelength, and $K_n(x)$ are the modified Bessel functions of the second kind. 

The spectral shape for the ULDM signal in the isotropic limit is shown in blue in the left panel of Figure~\ref{fig:spectra}. At frequencies smaller than $f\lesssim1/\tau$, the power spectrum grows like $S^{\scriptscriptstyle\rm DM}\propto 1/f^4$; while for $f\gtrsim1/\tau$ the power spectrum decays exponentially. In the right panel of the same figure, we compare the dipole structure of the isotropic ULDM signal with the Hellings \& Downs correlations expected for a gravitational wave signal. 

\subsection{Anisotropic Distribution}
The solar system revolution around the galactic center induces a \emph{dark matter wind} oriented in the opposite direction of the solar system velocity. Because of this effect, the DM velocity distribution in the solar system reference frame is not isotropic and is characterized by a non-zero mean velocity, $\vec{v}_0$:
\begin{equation}\label{eq:dm_v_dist}
    f(\vec{v}) = \frac{\rho_\oplus/m_\phi}{(2\pi\sigma^2)^{3/2}}
\exp\left[ - \frac{(\vec v-\vec v_0)^2}{2\sigma^2} \right]\,,
\end{equation}
where $\vec{v}_0=v_0\hat v_0$ is the mean DM velocity in the solar system reference frame. In this analysis, we will use $v_0=232\;{\rm km}/{\rm s}$ and $\hat v_0=\left[-0.47, 0.41, -0.78\right]$~\cite{1998gaas.book.....B, Sch_nrich_2010}.\footnote{Here we ignore the subleading effect induced by Earth's revolution around the Sun. The vector $\hat v_0$ is given in the equatorial coordinate system. }

Similarly, the orbital motion of pulsars will induce a DM wind in the pulsars' reference frame. Due to differences in the peculiar velocity, each pulsar will experience its own DM wind velocity. However, given that pulsars used in PTA observation are within a few kpc from the solar system over which rotational velocities are not expected to change substantially, we may assume that they all share the same wind velocity as long as the peculiar velocity of each pulsar is smaller than the DM wind velocity. This approximation will be used to derive the full PTA signal in the anisotropic limit, although it will not affect our final results as those will be derived only using the Earth term and neglecting the contribution from the pulsar term.

Assuming that the DM velocity distribution is given by Eq.~\eqref{eq:dm_v_dist} both in the Earth and pulsars' reference frame, the spectrum and the overlap reduction function are most conveniently written as 
\begin{equation}\label{eq:P_uldm_ani}
    \Gamma_{ab}^{\scriptscriptstyle\rm DM} S^{\scriptscriptstyle\rm DM}(f)
    = 
    \Gamma^\perp_{ab} S^\perp(f) + \Gamma^{\newparallel}_{ab} S^{\newparallel}(f)\,.
\end{equation}
The correlation functions are given by 
\begin{align}
    \Gamma^\perp_{ab} &=
    \frac{1}{2}(1+\delta_{ab}) 
    \Big[\hat n_a\cdot \hat n_b 
    -  (\hat n_a\cdot \hat v_0) (\hat n_b\cdot \hat v_0) \Big],
    \label{eq:orf_perp}
    \\
    \Gamma^{\newparallel}_{ab} &=
    \frac{1}{2}(1+\delta_{ab})  (\hat n_a\cdot \hat v_0) (\hat n_b\cdot \hat v_0)\,,
    \label{eq:orf_para}
\end{align}
where the term proportional to $\delta_{ab}$ is the pulsar term, while the remaining one gives the Earth term. In the anisotropic limit, inter-pulsar correlations for the ULDM signal are no longer a simple function of the pulsars' angular separation but depend on the positions of the pulsars in the sky. In the right panel of Figure~\ref{fig:spectra}, we report the value of $\Gamma_{ab}^A$ for all the pulsar pairs in our 15-year mock data set. It is clear from this figure that, for a fixed angular separation, there is a spread in the level of correlations due to different locations of the pulsars with respect to the orientation of the DM wind. The spectral functions appearing in Eq.~\eqref{eq:P_uldm_ani} can be written as 
\begin{equation}\label{eq:}
    S^A(f) = \frac{\bar a^2 \tau}{(2\pi f)^4} H^A(f, v_0)\,,
\end{equation}
for $A = \perp, \newparallel$. In the left panel of Figure~\ref{fig:spectra}, we compare the functions $S^A(f)$ (light and dark green lines) with the power spectrum in the isotropic limit. Detailed expressions and derivations for the $H^A$ functions are provided in Appendix~\ref{app:timing_residuals}.\footnote{The model files needed to implement the ULDM signals discussed in this section into the PTA analysis code \texttt{PTArcade}~\cite{andrea_mitridate_2023, Mitridate:2023oar} can be found at this link: \href{https://zenodo.org/records/10534322}{https://zenodo.org/records/10534322}.}

\section{Analysis}\label{sec:analysis}
In this section, we discuss noise and signal modeling, the data sets used in this analysis, and the statistical tools used to analyze them. 
\subsection{Noise Modeling and the PTA Likelihood}\label{subsec:pta_likelihood}
The statistical tools needed to model PTA timing residuals have been extensively discussed in the literature (see, e.g., Refs.~\cite{NANOGrav:2015aud, NANOGrav:2018hou}). In this section, we provide a short overview of these tools paying particular attention to how we can use them to model the ULDM signals derived in the previous section. 

The pulsar timing residuals receive contributions from several noise and astrophysical sources. All these sources are usually described by using a phenomenological model containing three main components: white noise, time-correlated stochastic processes (also known as red noise), and the impact of small errors in the fit to the timing-ephemeris parameters. Specifically, this phenomenological model parametrizes the timing residuals as:
\begin{equation}\label{eq:timing_residuals}
    \vec{\delta t} = \vec{n}+\vec{F}\,\vec{a}+\vec{M}\,\vec{\epsilon}\,,
\end{equation}
where $\vec{\delta t}$ is a vector containing all the $N_{\rm TOA}$ measured timing residuals. 

The first term on the right-hand side of Eq.~\eqref{eq:timing_residuals}, $\vec{n}$, describes the white noise that is assumed to be left in each of the timing residuals after subtracting all known systematics. White noise is assumed to be a zero mean normal random variable, fully characterized by its covariance. Assuming that a single instrument measures all the TOAs using a wideband approach, the white-noise covariance matrix reads:
\begin{equation}\label{eq:wn_cov}
\langle n_{a,i} n_{b,j}\rangle=\mathcal{F}_a^2\left[\sigma_{a,i}^2+\mathcal{Q}_a^2\right]\; \delta_{ij}\delta_{ab}\,,
\end{equation}
where $i$ and $j$ index the TOAs, $a$ and $b$ index pulsars, $\sigma_{a,i}$ is the TOA uncertainty for the $i$th observation, $\mathcal{F}$ is the {\it Extra FACtor} (EFAC) parameter, and $\mathcal{Q}$ is the {\it Extra QUADrature} (EQUAD) parameter.\footnote{For the analysis of the NANOGrav 12.5-year data, which was derived with narrowband observations, we also include correlated noise across radio frequencies as
\begin{align}
    \langle n_{a,i} n_{b,j}\rangle\supset\mathcal{J}^2\mathcal{U}_{ij}\,
\end{align}
where $\mathcal{J}^2$ is the ECORR parameter, and $\mathcal{U}_{ij}$ is a block diagonal matrix with unit entries for TOAs in the same observing epoch and zeros for all others entries.}

The second term on the right-hand side of Eq.~\eqref{eq:timing_residuals} describes time-correlated stochastic processes, which include pulsar-intrinsic red noise, and any stationary stochastic signal. These processes are modeled using a Fourier basis of frequencies $f_i\equiv i/T_{\textrm{obs}}$, where $i$ indexes the harmonics of the basis, and $T_{\textrm{obs}}$ is the timing baseline, extending from the first to the last recorded TOA in the complete PTA data set. Since we are generally interested in processes that exhibit long timescale correlations, this expansion is truncated after, $N_f$, frequency bins. In this work we will set $N_f=30$ for intrinsic red noise processes, and $N_f=14$ for GWB and ULDM signals in the mock and NANOGrav 12.5-year data sets.
This set of $N_f$ sine--cosine pairs evaluated at the different observation times are contained in the Fourier design matrix, $\vec{F}$. The Fourier coefficients of this expansion, $\vec{a}$, are assumed to be normally distributed random variables with zero mean and a covariance matrix, $\langle\vec{a}\vec{a}^{\textrm T}\rangle=\vec{\phi}$, given by
\begin{equation}
    \label{eq:red_cov}
    [\phi]_{(ai)(bj)}=\delta_{ij}
    \left( \delta_{ab}\varphi_{a,i}
    + \Gamma_{ab}^{\scriptscriptstyle\rm GW}\Phi_{i}^{\scriptscriptstyle\rm GW} 
    +\Gamma_{ab}^{\scriptscriptstyle\rm DM}\Phi_{i}^{\scriptscriptstyle\rm DM}\right)\,,
\end{equation}
where $a$ and $b$ index the pulsars, and $i$ and $j$ index the frequency harmonics.
The first term in Eq.~\eqref{eq:red_cov} describes intrinsic pulsar noise, and is parametrized as 
\begin{equation}
    \varphi_{a,i} = \frac{A_a^2}{12\pi^2}\frac{1}{T_{\rm obs}}\left(\frac{f_i}{{\rm yr}^{-1}}\right)^{\gamma_a}{\rm yr}^3\,.
\end{equation}
The second term in Eq~\eqref{eq:red_cov} describes the contribution to the timing residuals induced by a GWB. The ORF for this contribution is given by the well-known Hellings-Downs (HD) function~\cite{Hellings:1983fr}:
\begin{equation}
    \Gamma_{ab}^{\scriptscriptstyle\rm GW} = \frac{1}{2} \delta_{ab} + \frac{1}{2} - \frac{1}{4} x_{ab} + \frac{3}{2} x_{ab} \ln x_{ab}, 
\end{equation}
where $x_{ab} = ( 1- \hat n_a \cdot \hat n_b) /2$. In this analysis, we will parametrize the GWB common spectrum as 
\begin{equation}
    \Phi_i^{\scriptscriptstyle\rm GW}=\frac{A_{\scriptscriptstyle\rm GW}^2}{12\pi^2}\frac{1}{T_{\rm obs}}\left(\frac{f_i}{{\rm yr}^{-1}}\right)^{\gamma_{\scriptscriptstyle\rm GW}}{\rm yr}^3\,.
\end{equation}
Finally, the last term in Eq.~\eqref{eq:red_cov} describes the stochastic ULDM contribution to the timing residual and is related to the timing residuals' PSD by 
\begin{equation}
    \Gamma_{ab}^{\scriptscriptstyle\rm DM}\Phi_{i}^{\scriptscriptstyle\rm DM}=\Gamma_{ab}^{\scriptscriptstyle\rm DM}S^{\scriptscriptstyle\rm DM}(f_i)\Delta f\,,
\end{equation}
where $\Delta f=1/T_{\rm obs}$. In our analysis, we will always consider non-isotropic DM velocity distribution, such that $S_{ab}^{\scriptscriptstyle\rm DM}(f)$ is given by Eq.~\eqref{eq:P_uldm_ani}. However, since we are interested in probing the ULDM abundance within the solar system, we will neglect the pulsar term for the ULDM signal, which corresponds to neglecting the $\delta_{ab}$ terms in the ORFs given in Eqs.~\eqref{eq:orf_perp} and \eqref{eq:orf_para}.

The third and last term in Eq.~\eqref{eq:timing_residuals} describes the impact that linear deviations from the initial best-fit values for the $m$ timing-model parameters have on the timing residuals. The \emph{design matrix} $\vec{M}$ is a $N_{\rm TOA}\times m$ matrix, which contains the partial derivatives of the TOAs with respect to the timing-model parameters (evaluated at the best-fit values), and $\vec{\epsilon}$ is a vector containing the linear offset from the best-fit values of the timing model parameters.

Given this parametrization for the timing residuals, we can use the two-step marginalization of the timing and noise parameters described in Ref.~\cite{NANOGrav:2023icp} to obtain the PTA likelihood function
\begin{equation}\label{eq:pta_likelihood}
    p(\vec{\delta t}|\vec{\eta})=\frac{\exp\left(-\frac{1}{2}\vec{\delta t}^T\vec{K}^{-1}\vec{\delta t}\right)}{\sqrt{{\rm det}(2\pi\vec{K})}}\,,
\end{equation}
where $\vec{\eta}$ contains all the model parameters (i.e. $\mathcal{F}_a$ and $\mathcal{Q}_a$ plus all the red noise parameters), and $\vec{K}=\vec{D}+\vec{F}\vec{\phi}\vec{F}^T$. Here, $\vec{D}=\vec{N}+\vec{MEM}^T$, where $\vec{N}$ is a diagonal matrix whose non-zero elements are given by Eq.~\eqref{eq:wn_cov}, and $\vec{E}=\langle\vec{\epsilon\epsilon}^T\rangle$ is set to be a diagonal matrix of very large values ($10^{40}$), which effectively means that we assume flat priors for the parameters in $\vec{\epsilon}$.

\begin{table}[t]
\bgroup
\renewcommand{\arraystretch}{1.5}
\setlength\tabcolsep{6.pt}
{\footnotesize
	\begin{tabular}{ccc}
		\toprule
		\textbf{Parameter}	                       &   \textbf{Description}                 &   \textbf{Prior}   \\ \hline
		$\mathcal{F}_{a}$                          &   EFAC$^*$                             &   $[0.01,\,10]$    \\   
		$\log_{10}(\mathcal{Q}_{a}/{\rm s})$                 &   EQUAD$^*$                            &   $[-8.5,\,-5]$    \\ \midrule
		$\log_{10} A_a$                            &   intrinsic red-noise amplitude        &   $[-20,\,-11]$    \\
		$\gamma_a$                                 &   intrinsic red-noise spectral index   &   $[0,\,7]$        \\ \midrule
		$\log_{10} A_{\scriptscriptstyle \rm GW}$  &   GWB signal amplitude                 &   $[-20,\,-11]$    \\
		$\gamma_{\scriptscriptstyle \rm GW}$       &   GWB signal spectral index            &   $[0,\,7]$        \\ \midrule
		$\log_{10}(\rho_\oplus/\rho_0)$                   &   normalized local DM density          &   $[0,\,12]$       \\
		$\log_{10}(m_\phi/{\rm eV})$                    &   DM mass                              &   $[-19,\,-12]$    \\
		$\sigma\;[{\rm km/s}]$                                   &   DM velocity dispersion               &   $[150,\,250]$    \\
        \bottomrule
	\end{tabular}
}
\egroup
\caption{Prior distributions for the parameters used in this work. The parameters marked with a $^*$ were used only in single pulsar analyses.}
\label{tab:priors}
\end{table} 

\subsection{Setting Constraints and Projections}
The goal of our analysis is to set constraints and projections on the local ULDM density as a function of the ULDM mass. To do this, we make use of Bayes inference, a technique that uses Bayes’ rule of conditional probabilities to derive probability distribution for the parameters of the statistical model used to describe the data. In our case, these parameters are white noise parameters ($\mathcal{F}_a$ and $\mathcal{Q}_a$) and the ones contained in the covariance matrix defined in Eq.~\eqref{eq:red_cov}, which include intrinsic red noise parameters ($A_a$ and $\gamma_A$), GWB parameters ($A_{\rm GWB}$ and $\gamma_{\rm GWB}$), and ULDM parameters ($\rho_\oplus$, $m_\phi$, and $\sigma$). Given the PTA likelihood of Eq.~\eqref{eq:pta_likelihood}, we can use Bayes' theorem to get
\begin{equation}
    p(\vec{\eta}|\vec{\delta t})\propto p(\vec{\delta t}|\vec{\eta})p(\vec{\eta})
\end{equation}
$p(\vec{\eta}|\vec{\delta t})$ is the posterior probability distribution, $p(\vec{\delta t}|\vec{\eta})$ is the PTA likelihood function given in Eq.~\eqref{eq:pta_likelihood} and implemented using the \texttt{ENTERPRISE}~\cite{2019ascl.soft12015E} and \texttt{enterprise\_extensions}~\cite{enterprise} packages, and $p(\vec{\eta})$ are the prior probability distributions for the noise and ULDM parameters reported in Table~\ref{tab:priors}. 

Following the standard practice~\cite{EPTA:2021fqa, NANOGrav:2020bcs, Goncharov:2021oub, Perera:2019sca}, we first perform a single-pulsar noise analysis to derive the maximum likelihood values for each pulsars' white noise parameters. In this single-pulsar analysis, we model the timing residuals only using white and intrinsic red noise, such that the PTA likelihood will only depend on the parameters $\mathcal{F}_a$, $\mathcal{Q}_a$, $A_a$, and $\gamma_a$. GWB and ULDM parameters are not included, such that the covariance matrix in Eq.~\eqref{eq:red_cov} reads $\vec{\phi}=\vec{\varphi}$.

After this single-pulsar analysis, we performed a full PTA analysis in which the white noise parameters are fixed to their maximum likelihood value extracted from the single-pulsar analysis, while all the other parameters are varied within the prior ranges summarized in Table~\ref{tab:priors}.

Given the posterior distribution, we marginalize over all parameters except the local ULDM density, $\rho_\oplus$, and the ULDM mass $m_\phi$. The constraints on the local ULDM in each mass bin are then set to the $95^{\rm th}$ percentile of the ULDM density in that bin. 

Practically, we derive marginalized posterior distributions by using Markov Chain Monte Carlo (MCMC) techniques to sample randomly from the posterior distributions. To assess convergence of our MCMC runs we use the Gelman-Rubin statistic, $R$, and require $R< 1.02$ for all parameters including pulsar intrinsic red noise parameters. All the MCMC runs performed in this analysis used the \texttt{PTMCMC} sampler~\cite{justin_ellis_2017_1037579}.

\subsection{Data Sets}\label{subsec:mock}

\subsubsection{NANOGrav 12.5-year data set}
The NANOGrav 12.5-year data set~\cite{NANOGrav:2020gpb} consists of observations of 47 millisecond pulsars made between July 2004 and June 2017 by the Arecibo Observatory and the Green Bank Telescope. 
Most pulsars in this data set were observed approximately once per month, except six pulsars that were observed once per week as part of a high-cadence campaign carried out at the Green Bank Telescope since 2013 and at the Arecibo Observatory since 2015.
In our analysis, we will use the data from the 45 pulsars of this data set that have an observation baseline longer than 3 years. 
\subsubsection{Mock data sets}
We create our mock data using \texttt{libstempo}~\cite{2020ascl.soft02017V}, a Python wrapper of \texttt{TEMPO2}~\cite{Edwards:2006zg, Hobbs:2006cd}. We build the data set starting from a core catalog consisting of the 67 pulsars contained in the NANOGrav 15-year data set~\cite{NANOGrav:2023hde}, for which we assume an observation baseline of 15 years.\footnote{The NANOGrav 15-year data set contains in total of 68 pulsars. Our core catalog is based on the 67 that were used in the NANOGrav most recent GWB search~\cite{NANOGrav:2023gor}} This core catalog is expanded by adding 33 pulsars every 5 years until a total observing time of 30 years is reached. Shorter observing baselines are derived by slicing this data set into smaller catalogs. 

The final data set consists of 166 pulsars placed in random sky locations. Of these pulsars, 67 have an observing baseline of 30 years, while the remaining 99 are divided into three blocks which have a total observing baseline of 15, 10, and 5 years. For all pulsars, we assume an observing cadence of 3 weeks, with small random fluctuations added on top. Each TOA in the mock data set is associated with a TOA error, $\sigma_{a, i}$, which we derive by sampling from a normal distribution whose mean and standard deviation are reported in Table~\ref{tab:noise_dist}. The synthetic TOAs are injected with withe and red noise, plus a GWB signal. All these processes are described using the statistical models discussed in Section~\ref{subsec:pta_likelihood}. For each pulsar, the noise parameters are randomly sampled from the distributions given in Table~\ref{tab:noise_dist}, while the GWB parameters are set to $\log_{10}A_{\rm GWB}=-14.2$ and $\gamma_{\rm GWB}=3.2$.

\begin{table}[t]
\bgroup
\renewcommand{\arraystretch}{1.5}
\setlength\tabcolsep{3.pt}
{\footnotesize
	\begin{tabular}{ccc}
		\toprule
		\textbf{Parameter}	                       &   \textbf{Description}                 &   \textbf{Distribution}   \\ \hline
		$\sigma_{a,i}$                             &   TOA uncertainty                      &   $\mathcal{N}(400\,{\rm ns} , 200\,{\rm ns})$    \\ \midrule
		$\mathcal{F}_{a}$                          &   EFAC                                 &   $\mathcal{N}(1, 0.05)$    \\   
		$\log_{10}\mathcal{Q}_{a}$                 &   EQUAD                                &   $\mathcal{N}(-8.5, 1)$    \\ \midrule
		$\log_{10} A_a$                            &   intrinsic red-noise amplitude        &   $\mathcal{N}(-16, 1)$    \\
		$\gamma_a$                                 &   intrinsic red-noise spectral index   &   $\mathcal{U}(1, 5)$        \\ 
        \bottomrule
	\end{tabular}
}
\egroup
\caption{Probability distributions used to generate values of noise parameters.\label{tab:noise_dist} }
\end{table} 

\section{Results}\label{sec:result}

\subsection{Solar system dark matter density}\label{subsec:solar_den}
The results of our analysis are summarized in Fig.~\ref{fig:summary} where we show the constraints and projections on the average ULDM density obtained by analyzing the NANOGrav 12.5-year data set and a simulated data set with increasing observing baseline and number of pulsars. Full posterior distributions for the ULDM and GWB parameters are reported in Appendix~\ref{app:post}. 

The maximum sensitivity is achieved at the ULDM masse, $m_\phi\sim 1/(T_{\rm obs}\sigma^2)$, for which the ULDM coherence time approximately match the total observing baseline. At lower masses, the sensitivity weakens exponentially, while at higher masses they decrease like $\rho_\oplus\propto m_\phi^{3/2}$.
This is expected as for $m_\phi<1/(T_{\rm obs}\sigma^2)$ the ULDM signal within the PTA band is exponentially suppressed, while for $m_\phi>1/(T_{\rm obs}\sigma^2)$ the amplitude of the ULDM signal scales like $S^{\scriptscriptstyle \rm DM}\propto \rho_\oplus^2/m_\phi^3$, while the ULDM signals are entirely within the PTA frequency band.

It is important to stress that PTAs are sensitive to the DM abundance near the solar system, a quantity for which we have no direct experimental measure. Indeed, typical measurements of the DM abundance are derived using large-scale properties of the Milky Way, and probe volumes of order $\mathcal{O}(10^6\;{\rm pc}^3)$ or larger (for a review on measurements of the local DM abundance see for example Refs.~\cite{Read:2014qva, deSalas:2020hbh}). In other words, the result of these measurements, i.e. $\rho_0\sim0.4\;{\rm GeV/cm}^3$, do not exclude the possibility of DM overdensities at much smaller scales.

To the best of our knowledge, the strongest constraints on the dark matter density within the solar system come from solar system ephemerides, which constraint the DM density to be smaller than $\rho_\oplus / \rho_0 \lesssim 2\times 10^5$ at the Earth orbit, and $\rho_\oplus / \rho_0 \lesssim 2\times 10^4$ at the Mars/Saturn orbit~\cite{Pitjev:2013sfa}. Additional constraints arise from lunar laser ranging and LAGEOS geodetic satellite, $\rho_\oplus / \rho_0 \lesssim 10^{11}$~\cite{Adler:2008rq}, and the motion of asteroids in the solar system, $\rho_\oplus / \rho_0 \lesssim 6\times 10^6$~\cite{Tsai:2022jnv}. Our result shows that PTAs provide the strongest probe on ULDM densities around the solar system in the mass range of $m_\phi\sim10^{-18}\;{\rm eV}-10^{-16}\;{\rm eV}$.

\subsection{Dark matter substructures}
We perform a similar setup analysis while, this time, fixing the velocity dispersion to $\sigma = 10,\, 50,\, 164\,{\rm km/sec}$ and assuming vanishing mean velocity $v_0=0$. This analysis is motivated by the potential existence of dark matter substructures near the solar system with distinct kinematic properties. This may include a hypothetical dark disc~\cite{Read:2008fh}, stream-like structures accompanying stellar streams~\cite{Evans:2018bqy, OHare:2018trr, OHare:2019qxc}, or dense ULDM structures bound to the solar system via capture processes~\cite{Banerjee:2019epw, Banerjee:2019xuy, Budker:2023sex}. 
This particular analysis examines the capacity of PTA observations to probe cold dark matter substructures with small velocity dispersion.

\begin{figure}
\centering
\includegraphics{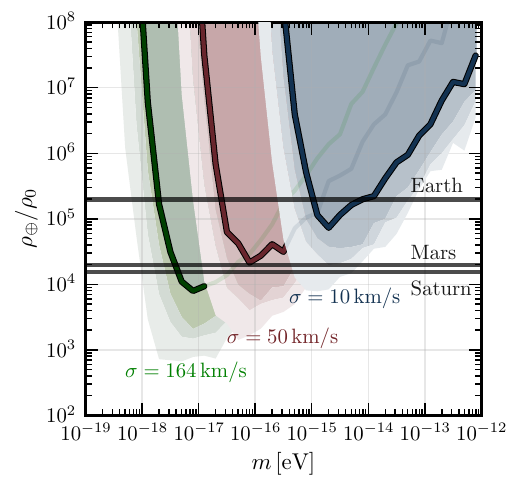}
\caption{Constraints and projections on the local ULDM abundance for cold DM substructures with $\sigma =10\,{\rm km/sec}$ (blue shaded regions), $50\,{\rm km/sec}$ (red shaded regions), and $164\,{\rm km/sec}$ (green shaded regions). The solid lines show the constraints derived by analyzing the NANOGrav 12.5-year data set. The other shaded regions show the constraints derived by analyzing the 15-year (dark-shaded), 20-year (medium-shaded), and 30-year (light-shaded) synthetic data sets.}
\label{fig:rho_cold}
\end{figure}

The result is shown in Figure~\ref{fig:rho_cold}. We observe two interesting features: (i) as the velocity dispersion decreases, the mass at which the strongest constraint appears shifts as $m_\phi \propto 1/ \sigma^2$ and (ii) the constraint at that mass scales as $\rho_\oplus \propto 1/ \sigma$. 
The first feature can be easily explained. From the analytic expression for the power spectrum \eqref{eq:P_uldm_iso}, one finds that the ULDM signal drops exponentially for $2\pi f \gtrsim m_\phi \sigma^2$. This suggests that the lowest mass that can be probed by PTA is $\hat{m}_\phi = 2\pi/(T_{\rm obs} \sigma^2)$, which explains the first feature shown in the figure.
The second feature can be explained similarly. The timing residual power spectrum induced by ULDM fluctuations is proportional to $S^{\scriptscriptstyle\rm DM}\propto \rho_\oplus^2 / m_\phi^3\sigma^4$. For $m_\phi = \hat m_\phi$, we then have $S^{\scriptscriptstyle\rm DM} \propto \rho_\oplus^2 \sigma^2$, which explains the $\rho_\oplus \propto 1/ \sigma$ scaling for the peak sensitivity. 

\section{Conclusion}\label{sec:conclusion}
We have considered the impacts of ULDM low-frequency stochastic fluctuations on pulsar timing observations and derived the overlap reduction function and the timing residual power spectrum induced by these fluctuations. 
We have found that these stochastic fluctuations allow us to probe ULDM in a mass range approximately six orders of magnitude higher than usual searches based on coherent ULDM oscillations.

To fully assess the prospect of PTA searches for this kind of fluctuations, we have analyzed (i) the NANOGrav 12.5-year data set and (ii) synthetic data with injected stochastic GWB and noise characteristics resembling actual PTA data sets. From the NANOGrav 12.5-year analysis, we have not found any signals of stochastic ULDM fluctuations, from which we place an upper bound on ULDM density near the solar system as $\rho_\oplus / \rho_0 \lesssim 4\times10^3$ at $m_\phi \simeq 10^{-17}\eV$. From the analyses of the simulated data set, we show that the sensitivity of future PTA observations could be improved up to an order of magnitude with the analysis of upcoming data sets. We have extended this analysis to probe cold dark matter substructures near the solar system and found that PTA will provide a strong probe of such objects for a wide range of mass. 

\begin{acknowledgments}
We thank the members of the NANOGrav timing working group for producing the NANOGrav 12.5-year data set used in this analysis, and the entire NANOGrav collaboration for useful comments on a preliminary form of this work. We also thank Xiao Xue for useful discussions. 
This work was supported by the Deutsche Forschungsgemeinschaft under Germany’s Excellence Strategy - EXC 2121 Quantum Universe - 390833306.
HK was also supported by the Munich Institute for Astro-, Particle and BioPhysics (MIAPbP) which is funded by the Deutsche Forschungsgemeinschaft (DFG, German Research Foundation) under Germany's Excellence Strategy – EXC-2094 – 390783311.
\end{acknowledgments}

\appendix
\section{ULDM fluctuations}\label{app:fluctuation}
In this appendix, following Ref.~\cite{Kim:2023pkx}, we provide a more detailed derivation of the power spectrum for density and pressure fluctuations of a ULDM scalar field, $\phi$, minimally coupled to gravity and whose action reads
\begin{equation}\label{eq:action}
    S = \int d^4x \sqrt{-g} \, \left[ \frac{1}{2} (\partial \phi)^2 - \frac{1}{2} m_\phi^2 \phi^2 \right]\,,
\end{equation}
where $g={\rm det}\,g_{\mu\nu}$ is the determinant of the metric $g_{\mu\nu}$.

The ULDM field can be modeled as the sum of many freely propagating plane waves~\cite{Kim:2021yyo}:\footnote{Here we are neglecting the $O(1\%)$ perturbations that the potential of the Sun can have on the ULDM wave functions.}
\begin{equation}
    \phi(t,\vec{x}) = \sum_{\vec k} \frac{1}{\sqrt{2m_\phi V}} 
    \left[
    \alpha_{\vec k} e^{-i k \cdot x}
    + \alpha_{\vec k}^* e^{+i k \cdot x}
    \right]
\label{phi_expansion}
\end{equation}
where the sum runs over the momenta of a particle quantized in a cubic box of volume $V$, and $(\alpha_{\vec k}, \alpha_{\vec k}^*)$ is a set of complex random number, whose underlying distribution is given by
\begin{equation}
    p(\alpha_{\vec k}) = \frac{1}{\pi f_{\vec k}}
    \exp\left[ - \frac{|\alpha_{\vec k}|^2}{f_{\vec k}} \right].
    \label{pdf}
\end{equation}
Here $f_{\vec k}$ is the mean occupation number, which will be later interpreted as a phase space distribution of dark matter particles. 
The above distribution does not depend on the phase of $\alpha_{\vec k}$; the phase follows the uniform distribution. Heuristic arguments for the distribution can be found in Refs.~\cite{Derevianko:2016vpm, Foster:2017hbq}. A field theoretical derivation is provided in Ref.~\cite{Kim:2021yyo}. 

Given this probability distribution, the ensemble average of any operator built from the $\phi$ field can be written as
\begin{equation}
\langle {\cal O} \rangle
= \int
\Big[ \prod_{\vec k} d^2\alpha_{\vec k} \, p(\alpha_{\vec k}) \Big] {\cal O}\,.
\label{avg_ens}
\end{equation}
We will now apply these results to study the density and pressure fluctuations of a ULDM field.


The stress-energy tensor for the ULDM field reads
\begin{equation}
    T_{\mu\nu}=\partial_\mu\phi\,\partial_\nu\phi-\frac{1}{2}g_{\mu\nu}\left[ (\partial\phi)^2-m_\phi^2\phi^2\right]\,,
\end{equation}
from which we can easily derive the ULDM energy density:
\begin{equation}
    \rho \equiv T_{00}= \frac{1}{2} 
    \Big[ (\dot\phi)^2 + (\nabla \phi)^2
    + m_\phi^2 \phi^2
    \Big].
\end{equation}
ULDM density fluctuations are defined as 
\begin{equation}
    \delta\rho = \rho - \langle \rho \rangle\,,
\end{equation}
with the mean energy density given by
\begin{equation}
    \langle\rho\rangle\simeq\frac{m_\phi}{V}\sum_{\vec k}f_{\vec k}\simeq m_\phi\int d^3{\vec v} f(\vec{v})
\end{equation}
where in the last step we have taken the continuum limit, changed the integration variable to the DM velocity, and adjusted accordingly the normalization of the phase space distribution.

Given the above results, we are now ready to derive the power spectrum of ULDM density fluctuations, $P_{\delta\rho}$, defined as 
\begin{equation}
    \langle\delta\rho(k)\delta\rho^*(k')\rangle\equiv(2\pi)^4\delta^{(4)}(k-k')P_{\delta\rho}(k)\,
\end{equation}
where we have defined
\begin{equation}
    \delta\rho(k)\equiv\int d^4x\,e^{-ik\cdot x}\,\delta\rho(x)\,.
\end{equation}
Using Eqs.~\eqref{phi_expansion}--\eqref{avg_ens} together with the relations
\begin{equation}
    \begin{split}
    \langle \alpha_{\vec k}^{\phantom{\ast}} \alpha_{\vec{k}'}^* \rangle &= f_{\vec k} \delta_{\vec{k}\vec{k}'} \\ 
    \langle \alpha_{\vec k}^{\phantom{\ast}} \alpha_{\vec q}^{\phantom{\ast}} \alpha_{\vec{k}'}^* \alpha_{\vec{q}'}^* \rangle &= f_{\vec k}^{\phantom{\ast}} f_{\vec q}^{\phantom{\ast}} (\delta^{\phantom{\ast}}_{\vec{k}\vec{k}'} \delta^{\phantom{\ast}}_{\vec{q}\vec{q}'} + \delta^{\phantom{\ast}}_{\vec{k}\vec{q}'} \delta^{\phantom{\ast}}_{\vec{q}\vec{k}'})\,,
    \end{split}
\end{equation}
we find that 
\begin{align}
P_{\delta\rho}(k)
\approx 
m_\phi^2 
\int d^3 v_1 d^3 v_2 \,
f(\vec v_1) f(\vec v_1)
(2\pi)^4 \delta(k - p_1 + p_2). 
\end{align}
Given the DM velocity distribution in the solar system:
\begin{equation}
f(\vec v) = \frac{\rho_\oplus/m_\phi}{(2\pi\sigma^2)^{3/2}}
\exp\left[
- \frac{(\vec v- \vec v_0)^2}{2\sigma^2}
\right],
\end{equation}
with $\vec v_0$ the mean dark matter velocity, $\rho_\oplus$ the mean DM density near the solar system, and $\sigma$ the velocity dispersion; the power spectrum for density perturbations takes the form
\begin{align}
P_{\delta\rho}(k)
= \frac{2\pi^2 \rho_\oplus^2}{m_\phi^4 \sigma^5}
\frac{\sigma}{v_k}
\exp\left[
- \frac{v_k^2}{4\sigma^2}
- \frac{\sigma^2}{v_k^2} \Big( \frac{\vec v_0 \cdot \vec v_k}{\sigma^2} - \omega\tau \Big)^2
\right]
\label{P_drho}
\end{align}
where $\vec v_k = \vec k / m_\phi$ and $\tau = 1 / m_\phi \sigma^2$.

\section{Timing residual}\label{app:timing_residuals}
In this Appendix we provide a detailed derivation of the ULDM signals discussed in Section~\ref{sec:ULDM}.

\subsection{Time delay}
We start by parametrizing the metric perturbations produced by ULDM fluctuations in terms of the scalar potentials $\Psi$ and $\Phi$:
\begin{equation}
ds^2 = (1 + 2 \Phi) dt^2 - (1 - 2 \Psi) dx^2. 
\end{equation}
Now, consider a bunch of photons emitted by a pulsar at space-time coordinates $x_{\rm em} = (t_{\rm em}, \vec{x}_p)$ and propagating to an observer on Earth that receives them at $x_{\rm obs}=(t_{\rm obs}, \vec{x}_e)$. The geodesic condition, $ds^2=0$, then requires that
\begin{equation}\label{eq:tobs}
    t_{\rm obs}=\Delta + t_{\rm em} -\int_{t_{\rm em}}^{t_{\rm obs}}dt\Big[\Phi(t,\vec{x}(t))+\Psi(t,\vec{x}(t))\Big]\,,
\end{equation}
where $\Delta = |\vec{x}_p - \vec{x}_e|$. Now consider a second bunch of photons emitted after a full rotation of the pulsar at $x'_{\rm em} = (t'_{\rm em}, \vec{x}'_p)$ and received at $x'_{\rm obs}=(t'_{\rm obs}, \vec{x}'_e)$. As before, we can use the geodesic condition to get 
\begin{equation}
    t_{\rm obs}'=\Delta' + t_{\rm em}' -\int_{t'_{\rm em}}^{t'_{\rm obs}}dt\Big[\Phi(t,\vec{x}'(t))+\Psi(t,\vec{x}'(t))\Big]
\end{equation}
where $\Delta' = |\vec{x}'_p-\vec{x}'_e|$. Assuming that the metric perturbations are small (i.e. $\Phi,\, \Psi \ll 1)$, we can approximate the integral on the right-hand side of this equation as 
\begin{equation}\label{eq:shap_int}
    \begin{split}
        &\int_{t'_{\rm em}}^{t'_{\rm obs}}dt\Big[\Phi(t,\vec{x}'(t))+\Psi(t,\vec{x}'(t))\Big]\\
        \simeq&\int_{t_{\rm em}+T_a}^{t_{\rm obs}+T_a}dt\Big[\Phi(t,\vec{x}'(t))+\Psi(t,\vec{x}'(t))\Big]\\
        \simeq&\int_{t_{\rm em}}^{t_{\rm obs}}dt\Big[\Phi(t+T_a,\vec{x}(t))+\Psi(t+T_a,\vec{x}(t))\Big]
    \end{split}
\end{equation}
where, in going from the first to the second line, we have used the approximation $t_{\rm em, obs}'\simeq t_{\rm em, obs}+T_a$, where $T_a$ is the rotational period of the $a$-th pulsar; while, in going from the second to the last line, we have redefined the integration variable as $t\to t+T_a$, and used that for the unperturbed photon trajectory, $\vec{x}(t)=\vec{x}_e+(t_{\rm obs}-t)\hat n_a$, we have that $\vec{x}'(t+T_a)\simeq\vec{x}(t)$. Therefore, using Eqs.~\eqref{eq:tobs}-\eqref{eq:shap_int} we can write 
\begin{widetext}
\begin{equation}\label{eq:dt_step_1}
    t'_{\rm obs}-t_{\rm obs}\simeq t'_{\rm em}-t_{\rm em} + \Delta'-\Delta +\int_{t_{\rm obs}-d_a}^{t_{\rm obs}}dt\Big[\Phi(t,\vec{x}(t))-\Phi(t+T_a,\vec{x}(t))+\Psi(t,\vec{x}(t))-\Psi(t+T_a,\vec{x}(t))\Big]\,,
\end{equation}
\end{widetext}
where in rewriting the integration limits we have used that $t_{\rm em}\simeq t_{\rm obs}-d_a$, where $d_a$ is the nominal distance between the observer and the $a$-th pulsar. The left-hand side of this equation is related  to the proper time interval between the detection of the two photon bunches for an observer on Earth as 
\begin{equation}\label{dtau}
    \Delta\tau \simeq\int_{t_{\rm obs}}^{t_{\rm obs}'}dt\Big[1+\Phi(t,\vec{x}_e)\Big]\simeq t'_{\rm obs}-t_{\rm obs}+T_a\Phi(t_{\rm obs},\vec{x}_e)\,
\end{equation}
where to evaluate the integral we have used that the scalar potential is approximately constant on the timescale of the pulsar period. In the rest of this section, we will discuss each contribution on the right-hand side of Eq.~\eqref{eq:dt_step_1}. 

The difference between the two emission times, $t'_{\rm em}-t_{\rm em}$, can be rewritten in terms of the pulsar period by requiring that the proper time interval between two emissions in the pulsar reference frame must equal the pulsar period:
\begin{equation}
    T_a\simeq\int_{t_{\rm em}}^{t_{\rm em}'}dt\Big[1+\Phi(t,\vec{x}_p)\Big]\simeq t'_{\rm em}-t_{\rm em} +T_a\Phi(t_{\rm em},\vec{x}_p)\,,
\end{equation}
where to evaluate the integral we have used that the scalar potential is approximately constant on the timescale of the pulsar period.

The difference $\Delta'-\Delta$ represents the change in path lengths the two consecutive photon bunches have to traverse. Indeed, fluctuations in the scalar field potential can accelerate the pulsar or the Earth and change their relative position:
\begin{equation}
    \frac{\Delta'-\Delta}{T_a}\simeq\hat n_a\cdot\int^{t_{\rm obs}}dt'\,\nabla\Big[\Phi(t',\vec{x}_e)-\Phi(t'-d_a,\vec{x}_p)\Big]
\end{equation}
where we have once again used that the scalar potential is approximately constant over a pulsar rotation and that $t_{\rm em}\simeq t_{\rm obs}-d_a$.

The integral in Eq.~\eqref{eq:dt_step_1} can be rewritten by expanding at first order the time dependence of the scalar potential:
\begin{equation}
    \begin{split}
        \Psi(t+T_a,\vec{x})&\simeq\Psi(t,\vec{x})+T_a\,\partial_t\Psi(t,\vec{x})\\
        \Phi(t+T_a,\vec{x})&\simeq\Phi(t,\vec{x})+T_a\,\partial_t\Phi(t,\vec{x})\,.
    \end{split}
\end{equation}
By substituting these expansion into Eq.~\eqref{eq:dt_step_1}, we can rewrite the integral as 
\begin{equation}
        -T_a\int_{t_{\rm obs}-d_a}^{t_{\rm obs}}dt\,\partial_t\Big[\Phi(t,\vec{x}(t))+\Psi(t,\vec{x}(t))\Big]\,.
\end{equation}
then using that $\partial_t=d/dt-\hat n_a\cdot\nabla$, we get
\begin{widetext}
\begin{equation}\label{shapiro_einstein}
    -T_a\Big[\Phi(t_{\rm obs},\vec{x}_e) - \Phi(t_{\rm obs}-d_a,\vec{x}_p)
        +\Psi(t_{\rm obs},\vec{x}_e)-\Psi(t_{\rm obs}-d_a,\vec{x}_p)\Big]
    +T_a\hat n_a\cdot\int_{t_{\rm obs}-d_a}^{t_{\rm obs}}dt \,\nabla\Big[\Phi(t,\vec{x}(t))+\Psi(t,\vec{x}(t))\Big]
\end{equation}
\end{widetext}

Substituting Eqs.~\eqref{dtau}--\eqref{shapiro_einstein} into Eq.~\eqref{eq:dt_step_1}, we find that for an observer on Earth, the interval between the arrival time of the two radiation pulses is given by 
\begin{equation}
    \frac{\Delta\tau}{T_a}\simeq 1+\left(\frac{\Delta T_a}{T_a}\right)_1+\left(\frac{\Delta T_a}{T_a}\right)_2+\left(\frac{\Delta T_a}{T_a}\right)_3
\end{equation}
where the three terms contributing to deviations in the observed pulsar's rotation period are given by
\begin{align}
\left(\frac{\Delta T_a}{T_a}\right)_1 &=
\Psi(t- d_a, \vec x_p) - \Psi(t, \vec x_e)
\\
\left(\frac{\Delta T_a}{T_a}\right)_2 &=
\hat n_a \cdot \int^t_{t-d_a} dt' \, 
\nabla \Big[
\Phi(t', \vec x(t')) + \Psi(t' , \vec x(t')) 
\Big]    
\\
\left(\frac{\Delta T_a}{T_a}\right)_3 &=
\hat n_a \cdot\int^t dt'\nabla\Big[\Phi(t',\vec{x}_e)-\Phi(t'-d_a,\vec{x}_p)\Big]
\end{align}
where in all these expressions we have replaced $t_{\rm obs}$ with $t$. The first two contributions come from the combination of Shapiro delay and proper time correction, while the last term is due to the Doppler shift of the arrival times caused by the Earth-pulsar relative motion. 

It is useful to rewrite these quantities in Fourier space:
\begin{align}
\left(\frac{\Delta T_a}{T_a}\right)_1 &= 
- \int \frac{d^4k}{(2\pi)^4} \, U(k, d_a, \hat n_a) e^{-i k \cdot x_{\rm obs}} 
\tilde \Psi(k)
\label{T1}\\
\left(\frac{\Delta T_a}{T_a}\right)_2 &=
- \int \frac{d^4k}{(2\pi)^4} \, 
\frac{\hat n_a \cdot \vec{k}}{\omega + \hat n_a \cdot {\vec k}} 
U(k, d_a, \hat n_a) e^{-i k \cdot x_{\rm obs}} 
\nonumber\\
&\times \big[ \tilde \Phi(k) + \tilde \Psi(k) \big]
\label{T2}\\
\left(\frac{\Delta T_a}{T_a}\right)_3 &= 
- \int \frac{d^4k}{(2\pi)^4} \, 
\frac{\hat n_a \cdot \vec{k}}{\omega} 
U(k, d_a, \hat n_a) e^{-i k \cdot x_{\rm obs}} \tilde \Phi(k)
\label{T3}
\end{align}
where $U(k, d_a , \hat n_a) = (1 - e^{i \omega d_a } e^{i \vec k \cdot \hat n_a d_a})$. Given that, as we have shown in Appendix~\ref{app:fluctuation}, for the dominant Fourier modes we have $\omega\lesssim m_\phi\sigma^2$ and $|\vec k|\lesssim m_\phi\sigma$, the Doppler effect, $(\Delta T_a/T_a)_3$, provides the dominant contribution. Note also that $\tilde \Phi \approx \tilde \Psi$ for low-frequency stochastic fluctuations. 

It is worthwhile to compare our result with Khmelnitsky and Rubakov~\cite{Khmelnitsky:2013lxt}. In their work, they have considered a coherently oscillating mode at $\omega =2m_\phi$. From \eqref{T1}--\eqref{T2}, it is clear that the coherently oscillating signal is strongest in $(\Delta T_a/T)_1$, while the other term is suppressed by the additional power of DM velocity. Using \eqref{T1}, it is straightforward to reproduce the results of Khmelnitsky and Rubakov as shown in the Appendix of Ref.~\cite{Kim:2023pkx}.

\subsection{Spectrum}
As discussed in the main text, the timing residual power spectrum induced by stochastic ULDM fluctuations can be written as 
\begin{align}
S^{\rm DM}_{ab}(f)
\approx
\frac{2G^2}{\pi^2 f^4} \!
\int \! \frac{d^3k}{(2\pi)^3} 
( \vec{k} \cdot \hat n_a) ( \vec{k} \cdot \hat n_b)
U_a U_b^*\frac{P_{\delta \rho}(f, \vec{k})}{k^4}\,,
\nonumber
\end{align}
where the density fluctuation power spectrum, $P_{\delta\rho}$, is given in \eqref{P_drho}.
For an explicit computation, we set the coordinate as
\begin{align}
\hat v_0 &=  (0,0,1),
\\
\hat k &= (\sin\theta \cos\phi, \sin\theta \sin\phi, \cos\theta),
\\
\hat n_a &= (\sin\theta_a \cos\phi_a, \sin\theta_a \sin\phi_a, \cos\theta_a),
\\
\hat n_b &= (\sin \theta_b \cos\phi_b, \sin\theta_b \sin\phi_b, \cos\theta_b).
\end{align}
After performing an explicit computation, one finds that the timing residual power spectrum can be written as a sum of two terms:
\begin{align}
S^{\scriptscriptstyle\textrm{DM}}(f) \Gamma_{ab}^{\scriptscriptstyle\textrm{DM}}(f)
= 
\Gamma^{\perp}_{ab} S^{\perp}(f)
+ \Gamma^{\newparallel}_{ab} S^{\newparallel}(f),
\end{align}
where the overlap functions are given in \eqref{eq:orf_perp}--\eqref{eq:orf_para}.
Each spectrum is given as
\begin{align}
S^A(f) &=
\frac{\bar a^2\tau}{(2\pi f)^4}
\left[
\frac{64}{\pi}
\int \frac{dx}{x} e^{-x^2/4}
C_A ( V_0, \bar \omega, x)
\right]
\\
&\equiv
\frac{\bar a^2\tau}{(2\pi f)^4}
H^A(f,V_0)\,,
\end{align}
where we have introduced $V_0 = v_0 / \sigma$ and $\bar\omega = \omega \tau$ and changed the integration variable to $x= v_k/\sigma$. 
Here the function $C_\perp$ and $C_{\newparallel}$ are given as
\begin{widetext}
\begin{align}
C_\perp ( V_0, \bar\omega, x) &=
+ \frac{1}{8V_0^3}
\Big[
\big(V_0 - \frac{\bar\omega}{x}\big) 
e^{-(V_0 + \frac{\bar\omega}{x})^2}
- \frac{\sqrt{\pi}}{2}
\Big(1 + 2 \Big(\frac{\bar\omega}{x}\Big)^2 - 2 V_0^2 \Big)
{\rm erf}(V_0 - \frac{\bar\omega}{x})
\Big]
+ (\bar\omega \to - \bar\omega), 
\\
C_{\newparallel} ( V_0, \bar\omega, x)&= 
- \frac{1}{4V_0^3}
\Big[
\big(V_0 - \frac{\bar\omega}{x}\big) 
e^{-(V_0 + \frac{\bar\omega}{x})^2}
- \frac{\sqrt{\pi}}{2} 
\Big( 1 + 2 \Big(\frac{\bar\omega}{x}\Big)^2 \Big) 
{\rm erf}(V_0 - \frac{\bar\omega}{x})
\Big]
+ (\bar\omega \to - \bar\omega),
\end{align}
\end{widetext}
In the limit $V_0\to 0$, one finds $H^\perp(f) = H^{\newparallel}(f) = (64/3\pi) K_0(\omega \tau)$, reproducing the results obtained in the isotropic limit. For the Bayesian analysis in the main text, we tabulate each function $H_A$ and use them for the numerical evaluations. 

\begin{figure*}[t]
\centering
\includegraphics[width=\textwidth]{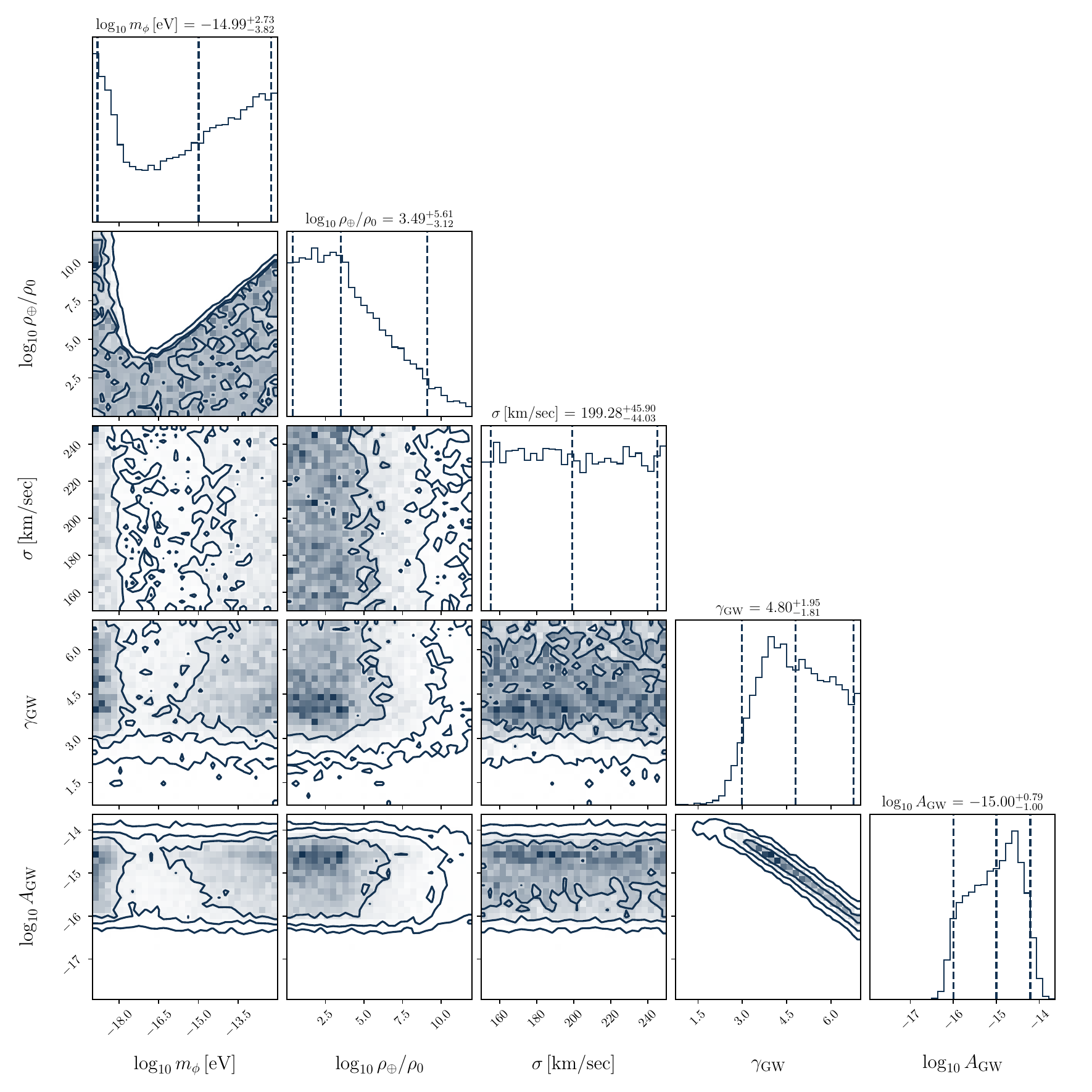}
\caption{The 2D posterior distribution for the analysis of NANOGrav 12.5-year data is shown. For a detailed description of the analysis, see Section~\ref{sec:analysis}. 
The quoted numbers above each panel correspond to the median and 90\% credible interval of each parameter, while each contour in the 2D histogram contains $68.3,\, 95.4$, and $99.7$\% of samples.
}
\label{fig:post_15}
\end{figure*}

\section{Posterior distribution}\label{app:post}

In Figure~\ref{fig:post_15}, we present the 2D posterior distribution for 5 parameters, 
$$(\log_{10}m_\phi, \, \log_{10} (\rho_\oplus/\rho_0), \, \sigma, \, \gamma_{\rm GW}, \, \log_{10}A_{\rm GW}),$$
obtained from the analysis of NANOGrav 12.5-year data. For detailed descriptions of the analysis, see Section~\ref{sec:analysis}. 

\newpage
\bibliography{ref}
\bibliographystyle{apsrev4-1}
\end{document}